\documentclass[preprint2]{aastex}

\newcommand{\etal}{et al.}

\newcommand{\kmsec}{\mbox{km s$^{-1}$}}

\def\gtorder{\mathrel{\raise.3ex\hbox{$>$}\mkern-14mu
             \lower0.6ex\hbox{$\sim$}}}

\def\ltsima{$\; \buildrel < \over \sim \;$}
\def\simlt{\lower.5ex\hbox{\ltsima}}
\def\gtsima{$\; \buildrel > \over \sim \;$}
\def\simgt{\lower.5ex\hbox{\gtsima}}

\begin{document}


\title{Photometric Typing Analyses of Three Young Supernovae with the
       Robotic Palomar 60-Inch Telescope} 


\author{Anne M. Rajala, 
        Derek B. Fox, 
        Avishay Gal-Yam\footnote{Hubble Fellow}, 
        Douglas C. Leonard\footnote{NSF Astronomy and Astrophysics
        Postdoctoral Fellow},
        Alicia M. Soderberg, 
        S.~R. Kulkarni}
\affil{Astronomy Department, MS 105-24, California Institute of
       Technology, Pasadena, CA 91125}
\email{amr@astro.caltech.edu, derekfox@astro.caltech.edu,
       avishay@astro.caltech.edu, leonard@astro.caltech.edu,
       ams@astro.caltech.edu, srk@astro.caltech.edu} 

\author{S. Bradley Cenko, 
        Dae-Sik Moon\footnote{Robert A. Millikan Fellow},
        Fiona A. Harrison}
\affil{Physics Department and Space Radiation Laboratory, MS
       220-47, California Institute of Technology, Pasadena, CA
       91125}
\email{cenko@srl.caltech.edu, moon@srl.caltech.edu, 
       fiona@srl.caltech.edu}





\begin{abstract}

We present photometric typing analyses of three young supernovae
observed with the Robotic 60-inch Telescope at Palomar Observatory
(P60).  This represents the first time that such photo-typing has been
attempted, in a blind fashion, on newly-discovered supernovae.  For
one of the target supernovae, SN\,2004cs, our photometry provided the
first constraint on the SN type, which we predicted would be type~Ia.
To the contrary, however, our subsequent Keck spectroscopy shows it to
be an unusual type~II supernova.  For each of the other two
supernovae, SN\,2004dh (type~II) and SN\,2004dk (type~Ib), our
photo-typing results are consistent with the known type of the event
as determined from ground-based spectroscopy; however, the colors of
SN\,2004dk are also consistent with a type~Ic or type~II
classification.  We discuss our approach to the challenges of
photo-typing -- contamination by host galaxy light and the unknown
photometric quality of the data -- in the case where it is desirable
to complete the analysis with just one night of observations.  The
growing interest in the properties and behavior of very young
supernovae, and the increased discovery rate for such events, mean
that prompt photo-typing analyses can provide useful input to
observational campaigns.  Our results demonstrate the value and
feasibility of such a project for P60, at the same time illustrating
its chief inherent shortcoming: an inability to identify new and
unusual events as such without later spectroscopic observations.

\end{abstract}





\section{Introduction}
\label{sec:intro}

Supernovae (SNe) are classified into several commonly-recognized
categories on the basis of features in their optical spectra.  These
categories -- the SN types Ia, Ib, Ic, and II -- are not merely
phenomenological distinctions, but reflect real differences in the
nature of the progenitor and the subsequent explosion (for a recent
review, see Filippenko 1997).

As understanding of the nature of these differences has grown -- and
moreover, as the number of supernova discoveries has increased to more
than 100 per year -- the SN follow-up community has become
increasingly specialized.  Type~Ia SNe have proven to be useful
standard candles for cosmography (Perlmutter et al.\ 1997; Riess et
al.\ 1998; Tonry \etal\ 2003; Knop \etal\ 2003; Riess \etal\ 2004),
and provided the first strong evidence for a cosmological constant or
quintessence.  Three type~Ic supernovae have been found in association
with gamma-ray bursts (SN\,1998bw: Galama \etal\ 1998; SN\,2003dh:
Stanek \etal\ 2003; Hjorth \etal\ 2003; Matheson \etal\ 2003;
SN\,2003lw: Cobb \etal\ 2004; Thomsen \etal\ 2004; Malesani \etal\
2004; Gal-Yam \etal\ 2004b), and indeed, type Ib and Ic supernovae
provide the most likely population of ``collapsar'' events (Woosley
1993; MacFadyen \& Woosley 1999).  Finally, interest in type~II events
has been focused by the type~II SN\,1987A and its associated neutrino
burst (Arnett \etal\ 1989 and references therein), which provided the
first direct observation of core collapse.  Type~II SNe may also prove
to be useful distance indicators, if indications that their
plateau-phase luminosities and expansion velocities are strongly
correlated (Hamuy \& Pinto 2002) prove to be correct.

Given the diversity of causes towards which SN observations may be
applied -- and the various strategies employed by groups that study
SNe -- it follows that the efficiency of follow-up observations for
any SN is dependent on the rapid identification of its type.  These
types are ultimately determined by analysis of the optical spectra.
However, spectroscopy time remains a somewhat precious commodity, and
there is often a significant time delay, of days to weeks, before a SN
can be properly typed in this fashion.

In two recent papers, Poznanski \etal\ (2002) and Gal-Yam \etal\
(2004a) have presented the outline of an alternative approach that can
provide a probabilistic estimate of the type of newly-discovered,
young supernovae using multi-color photometry.  With the proper
facilities this approach could be implemented on a rapid basis
following the discovery of each new SN.  Depending on the quality of
the data obtained, and assuming the supernova spectra used to develop
their test are sufficiently generic, the type of the SN can be
determined in this way with a high degree of confidence before any
spectra are taken.

The Poznanski-Gal-Yam method (hereafter ``PGM'') makes use of three-
or four-filter photometry of the supernova, with the several
magnitudes applied to calculate two colors for the object.  The
location of the object in this two-color space is then compared to the
locations of supernovae of various types and ages by reference to an
extensive spectroscopic library.  Since PGM works from the basis of
flux-calibrated optical (and to an extent, UV and Near-IR) spectra,
the specific choice of filters and acceptable range of SN redshifts is
not constrained a priori.  In fact, Poznanski and Gal-Yam have
implemented the method via an interactive web page%
\footnotemark\footnotetext{\url{http://wise-obs.tau.ac.il/$\sim$dovip/typing}},
their ``Typing Machine,'' which accepts input of an arbitrary set of
Sloan or Johnson-Cousins filters.  As Gal-Yam \etal\ (2004a) have
pointed out, the similar spectra of all young supernovae at redder
wavelengths ($\lambda>5500$\AA) means that classification of
low-redshift events is best done with measurements in the $B$, $g$,
$V$ and $R$ filters.

When the redshift of the supernova is known -- typically, because the
SN has been found in association with a catalogued local galaxy -- the
only remaining unknown is the quantity of extinction in the SN host
galaxy.  This parameter cannot be directly estimated from the
photometric data; instead, an ``arrow'' in the two-color space shows
the color correction that would account for one magnitude of
extinction ($A_V = 1$~mag) in the host.  Since this is more or less
the greatest extinction we expect to observe for supernovae discovered
in unfiltered optical searches, this provides a reasonable indication
of the associated systematic uncertainty.

The first application of PGM photo-typing by Poznanski \etal\ (2002)
evaluated the likely type of a supernova discovered during the course
of the Sloan Digital Sky Survey, SN\,2001fg, which had its redshift
and type determined by follow-up Keck spectroscopy (Filippenko \&
Chornock 2001).  Although the photometric analysis (based on the Sloan
discovery photometry; Vanden Berk \etal\ 2001) suggested a
$\simgt$1\,month-old type~Ia identification for the event, in accord
with the spectroscopy, this conclusion used the redshift of the event
as derived from the Keck data.  Thus it was more suggestive than
diagnostic of the utility of the PGM.

We are therefore interested in making the first observational tests of
PGM photo-typing, by obtaining four-filter imaging of young,
low-redshift supernovae.  We are in a unique position to perform these
tests, having recently completed the roboticization of the Palomar
60-inch (1.5-m) telescope (P60).  Approximately 80\% of the observing
time on P60 is devoted to transient astronomy, and in the absence of
high-priority campaigns we are able to obtain images of new SNe
quickly, within a day or two of the SN discovery.  Indeed, in the
2004B semester we have begun a Palomar Large Program, the ``Caltech
Core Collapse Program'' (CCCP; Gal-Yam \etal\ 2004c) to monitor the
photometric and spectroscopic evolution of a complete sample of
low-redshift core-collapse SNe over the course of a year.  Early-time
SN photo-typing observations make a natural supplement to these
extended CCCP photometric campaigns.

The organization of our paper is as follows.  In section 2, we review
our procedure for performing accurate photometry on the supernovae.
Section 3 presents our observations and data reduction and analysis.
In section 4, we photo-type our three target SNe, SN\,2004cs,
SN\,2004dh, and SN\,2004dk.  In section 5, we present our
spectroscopic observations of SN\,2004cs, demonstrating its unusual
type~II nature, and reflect on the erroneous photo-typing result for
this event.  Section 6 presents our conclusions and discusses future
prospects for this work.


\section{Methodology}
\label{sec:method}

Accurate photometry of a newly-discovered supernova presents at least
two challenges.  First, the SN will be superposed on some unknown
quantity of background light due to its host galaxy; and second, the
immediate field of the SN will have to be photometrically calibrated.
Moreover, if a quick turn-around is desired, this calibration will
generally have to be carried out in less than ideal conditions. 

We address the first of these issues by performing point-spread
function photometry on the SN and on several reference (PSF) stars,
using the DAOPHOT package (Stetson 1987, 1990) in the
IRAF\footnote{NOAO's Image Reduction and Analysis Facility,
\url{http://iraf.noao.edu}} environment.  Although this will fail to
account for any point-like component of the host galaxy light (e.g., a
compact underlying HII region), such contamination cannot be avoided
without reference pre- or post-supernova imaging; moreover, since we
are studying low-redshift events in resolved galaxies, the strength of
such contamination is not expected to be severe.

If the night is (close to) photometric, standard fields are analyzed
with the Stetson Photometric
Catalog\footnote{\url{http://cadcwww.dao.nrc.ca/standards/}}, using
the PHOTCAL package in IRAF to derive the photometric transformation
equations for the night, and apply these equations to the aperture
magnitudes of the PSF stars in the SN field.  The resulting ``true''
magnitudes of the PSF stars are then compared to their PSF magnitudes
to derive the PSF-to-aperture zero-point shift to apply to the
supernova PSF magnitudes.  If the night is not sufficiently
photometric, we retain the photo-typing observation sequence in the
P60 queue for roughly a week, and use the best available night for our
analysis.

In general, we aim to perform our analysis of each supernova promptly.
As such, we sacrifice some accuracy in attempting to calibrate the
images on the same night as the observations whenever possible instead
of waiting for a photometric night (which are rare at Palomar).  We
examine the coefficients of the transformation equations produced by
the standard fields to determine if the night is photometric.  If
these parameters are unreasonable (implying, for example, that stars
are appearing brighter at increased airmass), some subset of the
standard observations may be dropped.  To force reasonable
coefficients, the airmass term can be dropped from the fit entirely
and only standards taken at an airmass similar to the target field's
used to calculate the photometric and color shifts.  To control for
associated systematic effects -- e.g.\ clouds -- we can observe the SN
target field twice during the night.  This does not add much time to
the run, but allows the full analysis to be performed twice on
quasi-independent sets of data.  If consistent answers cannot be
obtained for the magnitudes of the reference stars across the two
observations, we may find that we can derive consistent answers for
their colors (i.e., to the extent that we are subject to varying
levels of approximately gray extinction).  This introduces an
uncontrolled systematic error, in that it assumes the SN is similar
enough in color to our reference stars that atmospheric and extinction
effects are comparable; however, it does at least allow for a
photo-typing measurement to be made.

The Stetson Standard Fields are not calibrated in the $g$ band; we
calculate $g$-band magnitudes for stars in the standard fields from
their $B$ and $V$ magnitudes using the transformation equations of
Smith \etal\ (2002).

To check the accuracy of the photometric errors estimated by IRAF, the
calculated magnitudes of the reference stars from two independent
data-sets are then compared.  Ideally, the second data-set is taken on
the same night, but it can also come from a different night with a
distinct set of transformation equations.  A chi-squared comparison
can then be made to reject outliers and estimate the factor by which
to inflate the IRAF errors to get a reduced $\chi^2$ value of one.  We
have found that this is generally necessary since IRAF tends to quote
errors that are too small, partly because they are derived from an
idealized model, and partly because the entire photometric calibration
process presumes ideal photometric conditions.  Separately, at these
beginning stages of the P60 our observations were subject to
significant seeing variations over the course of the night; the seeing
at P60 tending to be poor at the beginning of the night, then
improving and stabilizing by midnight.  A similar $\chi^2$ check is
done for the calculated aperture-to-PSF zero-point shifts, on a
star-by-star basis.  The derived multiplicative factors required to
make the IRAF-quoted errors acceptable are then applied to the SN
errors.

While one goal of this project is to create as automated a process as
possible, a minimal amount of human interaction helps to ensure robust
and repeatable results.  First, it is advisable to have a human
selecting or reviewing the reference stars selected in the SN field;
this does not take much time but is more reliable than IRAF's
selection methods, which occasionally select saturated stars, stars
near bad pixels, cosmic rays, or non-isolated stars.  Additionally,
constructing photometric calibrations for non-photometric nights can
lead to variations which make it difficult to trust a single blind fit
of the transformation equations.  Indeed, the three-step process of
progressively more conservative (and less precise) approaches to
photometric calibration that we describe above can probably only be
carried out in an interactive fashion, although an automated pipeline
could take the alternative approach of carrying out all three
approaches in parallel (later allowing human interaction to select the
best of the three).


\section{Observations and Data Reduction}
\label{sec:obs}

We selected candidates that were discovered at a provably young age
and were observable from Palomar Observatory.  The target supernovae
were observed by P60 on the first available night in $BgV\!R$ or
$BV\!RI$ (Table~\ref{tab:mags}).

All of our images were taken with the new camera at P60, which is a
two-amplifier, 2048$\times$2048 SITe CCD with a 12.9\arcmin\ field of
view at Cassegrain focus.  Readout (25~s for the full chip in
single-pixel binning mode) is accomplished by a two-channel Leach III
controller card; details of the new camera and robotic observing
system will be presented in a forthcoming paper.  Data were
overscan-subtracted, demosaicked, bias-subtracted, flat-fielded (using
dome flats taken the preceding afternoon), bad pixel-masked, and had a
blind-pointing world coordinate system (WCS) applied by the P60 image
analysis pipeline.  We refined this WCS by referencing Digitized Sky
Survey images of the field.

SN\,2004cs in UGC~11001 was discovered before peak (rising by $>$1~mag
in two days) by Li \& Singer (2004) on June 23.42 UT.  We observed the
SN on the night following discovery, June 24, 2004, starting at 06:59
UT.  Images were taken in the $B$, $g$, $V$ and $R$ filters at an
airmass of 1.06 with seeing of 1.8\arcsec.  Each of these exposures
lasted for 300 seconds.

This night was not photometrically calibrated, so additional images
were taken of the SN\,2004cs field, along with images of PG~1657, on
the night of July 30, beginning at 04:13 UT (1.8\arcsec\ seeing),
06:18 UT (1.7\arcsec\ seeing), and 08:30 UT (2.0\arcsec\ seeing).
These 30-second exposures were taken at three different airmasses,
ranging from 1.1 to 2.3.  PG~1657 is one of the Stetson Photometric
Standard Fields.  These standard fields were used to fit
transformation equations to the data in the $B$, $V$, and $R$ filters.
These parameters are listed in Table~\ref{tab:calib}.  Since these
terms are reasonable, we applied them to the reference stars in the
field of SN\,2004cs to derive their true magnitudes.  The two sets of
observations used to calculate the true magnitudes of the reference
stars were taken this night beginning at 04:23 UT (at 1.07 airmass and
1.9\arcsec\ seeing) and 07:34 UT (at 1.30 airmass and 1.6\arcsec\
seeing).  Each exposure was 60 seconds long.

In the case of SN\,2004cs, since calibration was done on a separate
night, the shifts of the reference stars were calculated separately;
the first by finding the average shift between the true magnitudes
(derived from the second night's data) and the instrumental aperture
magnitudes (from the first night's data), and the second shift by
finding the difference between the instrumental aperture magnitudes
and the PSF magnitudes on the first night.  These two shifts, applied
to the SN to get its true magnitudes (see Table 2), were respectively
$-$1.92(2)~mag and 0.0016(50)~mag in $B$, $-$1.83(4)~mag and
$-$0.76(2)~mag in $g$, $-$1.80(4)~mag and $-$1.06(3)~mag in $V$, and
$-$1.87(5)~mag and $-$1.4(6)~mag in $R$.

SN\,2004dh in MGC~+4-1-48 was discovered by Moore \& Li (2004) in an
image taken on July 21.47, and confirmed from an earlier image on July
11.45.  Prompt spectroscopy by Matheson \etal\ (2004) classified this
SN as a type~II before our analysis was complete.  We observed the SN
on 25 July 2004 beginning at 07:58 UT.  A $g$ band image was not
taken, and our analysis was performed with the $B$, $V$, $R$ and $I$
filters.  All exposures were taken for 120 seconds.  This set was
imaged at an airmass of 1.57 with 1.5\arcsec\ seeing.

For photometric calibration purposes, images were taken at a similar
airmass (1.597) at a similar time (beginning at 07:44 UT) of the
standard field PG~1657 in the $B$, $V$, $R$ and $I$ filters.  Each of
these exposures were taken for 30 seconds; at this time the seeing at
P60 was 1.4\arcsec.  Only one set of observations was used to
determine the photometric zero-points.  Since these images were taken
at nearly the same airmass and time, the airmass terms were fixed at
zero in the transformation equations.  The coefficients of these
equations are given in Table~\ref{tab:calib}.

The average zero-point shift was then calculated between the PSF
magnitudes of each reference star and its true magnitudes, to be
applied to the supernova's magnitudes (Table~\ref{tab:mags}).  These
shifts were $-$2.09(4)~mag in $B$, $-$2.71(1)~mag in $V$,
$-$3.03(2)~mag in $R$, and $-$3.36(3)~mag in $I$.

SN\,2004dk in NGC~6118 was discovered before peak by Graham \& Li
(2004) on August 1.19 UT.  This SN was identified as a type~Ic (Patat
\etal\ 2004b) before our analysis was complete; subsequent observations
of the event suggest it is more likely a type~Ib (Filippenko \etal\
2004).  We observed the SN on 3 August 2004 beginning at 04:02 UT, two
nights after its discovery, in the $B$, $g$, $V$ and $R$ filters.
Ninety-second eposures were taken in each filter at an airmass of
about 1.26 with seeing of 3.8\arcsec.

On this night, images of the standard field PG~1657 were also taken,
beginning at 05:56 UT, to photometrically calibrate the fields.  These
were taken in each filter ($B$, $V$, and $R$) for 30 seconds, at an
airmass of about 1.26 (similar enough to the airmass at which the SN
images were taken to fit transformation equations without an airmass
term).  These images were taken with seeing of about 1.65\arcsec.  The
transformation equations are given in Table~\ref{tab:calib}.

The true photometry of SN\,2004dk (see Table~\ref{tab:mags}) was then
derived by applying the same shifts as are calculated between the true
magnitudes and the PSF magnitudes of the reference stars.  These
shifts are $-$2.41(5)~mag in $B$, $-$2.85(2)~mag in $g$,
$-$3.20(1)~mag in $V$, and $-$3.70(1)~mag in $R$.


\section{Photo-typing of SN\,2004cs, SN\,2004dh and SN\,2004dk}
\label{sec:sntype}

\begin{figure*}
\includegraphics[width=17cm]{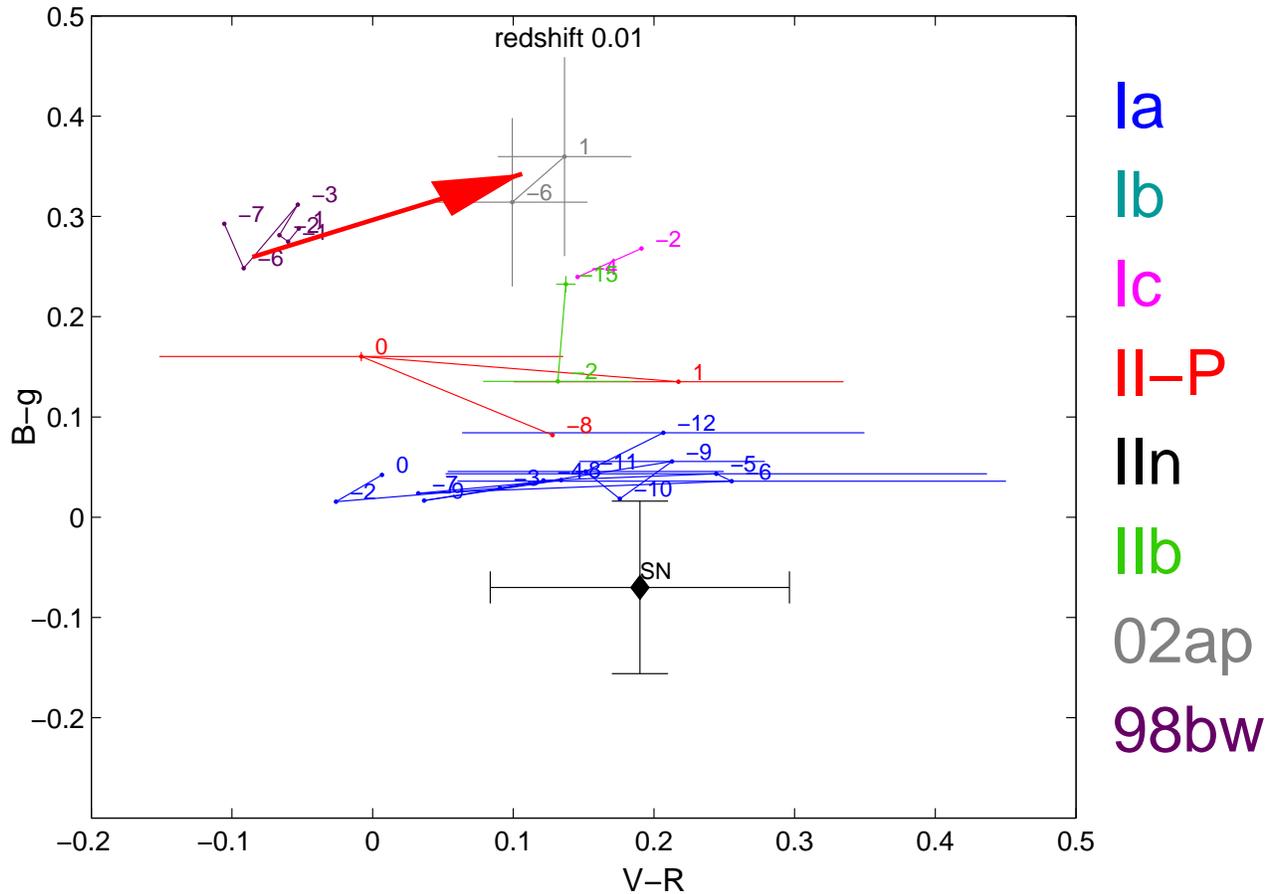}
\caption{$B-g$ and $V-R$ colors of SN\,2004cs on 2004 June 24.3 UT
(black diamond) compared to tracks for typical supernovae of each type
(and two distinctive individual events, SN\,1998bw and SN\,2002ap) for
SN ages from 20 days before to 1 day after maximum light.  The tracks
show the evolution in SN color for each of the comparison events, with
numbers giving the age in days relative to maximum light at each
point.  A young type~Ia identification for SN\,2004cs is indicated, as
we reported in Rajala \etal\ (2004a,b,c); however, our subsequent
spectroscopy has revealed the SN to be an unusual type~II event
(\S\protect\ref{sec:spec}).  The redshift of UGC~11001, the SN host
galaxy, is 0.014060.  The large red arrow indicates the color
correction that would be required if SN\,2004cs were subject to
$A_V=1$~mag of extinction in its host galaxy.  This plot was created
with the online PGM Typing Machine; see text for details.}
\label{fig:sncs}
\end{figure*}

\begin{figure*}
\includegraphics[width=17cm]{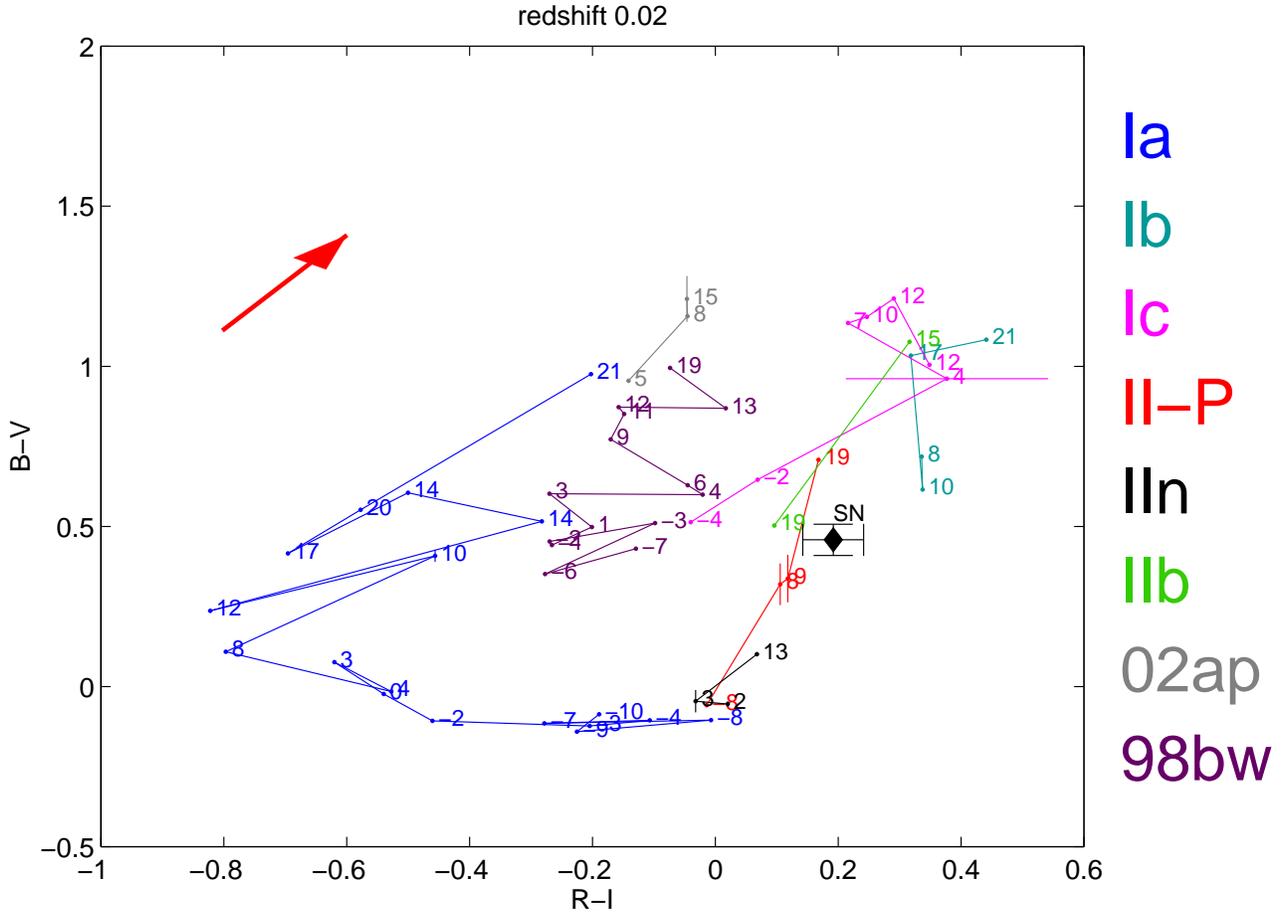}
\caption{$B-V$ and $R-I$ colors of SN\,2004dh on 2004 July 25.3 UT
(black diamond) compared to tracks of typical supernovae various types
(and two distinctive individual events) for SN ages from 20 days
before maximum light to 21 days after maximum light.  A type~II
identification for SN\,2004dh is indicated, in agreement with the
spectroscopic determination by Matheson \etal\ (2004).  The redshift
of MGC~+4$-$1-48, the SN host galaxy, is 0.019327.  This plot was
created with the online PGM Typing Machine; for description of the
figure elements, see Fig.~\protect\ref{fig:sncs}.  }
\label{fig:sndh}
\end{figure*}

\begin{figure*}
\includegraphics[width=17cm]{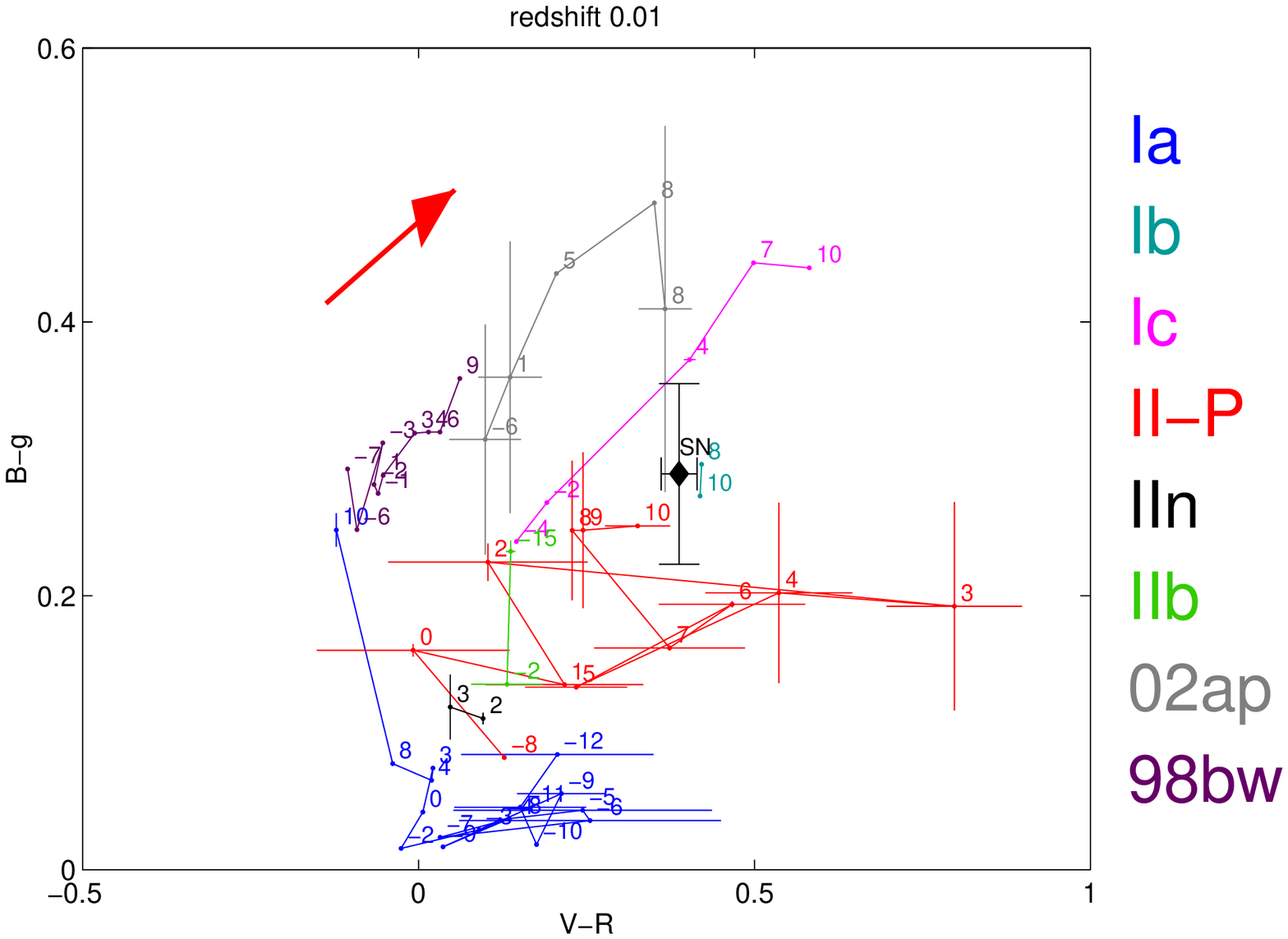}
\caption{$B-g$ and $V-R$ colors of SN\,2004dk on 2004 August 3.2 UT
(black diamond) compared to tracks of typical supernovae of various
types (and two distinctive individual events) for SN ages from 20 days
before maximum light to 10 days after maximum light.  A type~Ib,
type~Ic or type~II identification for SN\,2004dk is indicated.  The
type~Ib spectroscopic determination of Filippenko \etal\ (2004) is
consistent with these findings.  The redshift of NGC~6118, the SN host
galaxy, is 0.005247.  This plot was created with the online PGM Typing
Machine; for description of the figure elements, see
Fig.~\protect\ref{fig:sncs}.}
\label{fig:sndk}
\end{figure*}

To photo-type each of our three target supernovae, we used the
calibrated magnitudes (Table~\ref{tab:mags}), a restriction on the
likely age of the SN relative to maximum light, and the known redshift
of the SN host galaxy as input to the PGM Typing Machine.  The SN age
restriction is derived from the time of the last pre-discovery
observation, included in the SN discovery reports; for inclusion in
our ``young'' target sample, we required that the reference imaging be
taken within one month prior to discovery.  The resulting PGM Typing
Machine plots (Figures \ref{fig:sncs}, \ref{fig:sndh}, \ref{fig:sndk})
compare the colors of each SN with the colors of typical SNe of each
type, as well as the colors of several individual events of particular
notoriety: SN\,1998bw and SN\,2002ap (see Gal-Yam \etal\ 2004a).

For SN\,2004cs, the SN age was restricted to the range from 20~days
before to 1~day after maximum light, and the redshift of its host
galaxy, UGC~11001, is 0.014060.  The corresponding PGM plot
(Fig.~\ref{fig:sncs}) implies that this is a likely type~Ia event.  As
no one had yet announced the type of SN\,2004cs at the completion of
our initial analysis, we announced this result as Astronomer's
Telegram \#320 and IAU Circulars 8386 and 8387 (Rajala, Fox \& Gal-Yam
2004a,b,c).  Our subsequent spectroscopic observations of this SN,
however, reveal it to be an unusual type~IIb event; see
\S\ref{sec:spec} below.  

For SN\,2004dh, the SN age was restricted to the range from 20 days
before to 21 days after maximum light, and the redshift of its host
galaxy MGC~+4-1-48 is 0.019327.  The plot of SN\,2004dh
(Fig.~\ref{fig:sndh}) implies that it is a type~II, and quite likely
of sub-type II-P.  This agrees with the spectroscopic determination of
Matheson \etal\ (2004).

For SN\,2004dk, the SN age was restricted to the range from 20 days
before to 10 days after maximum light, and the redshift is that of its
host galaxy NGC~6118, 0.005247.  The plot of SN\,2004dk
(Fig.~\ref{fig:sndk}), indicates a core-collapse (e.g., non-Ia)
identification, although the data and template sets do not allow a
strong conclusion to be reached with regards to the specific type,
which may be type Ib, Ic or II-P.  This agrees with the type~Ib
identification for this event determined by Filippenko \etal\ (2004).
Alternatively, it also agrees with the earlier suggested type~Ic
identification of Patat \etal\ (2004b).


\section{Spectroscopic Typing of SN\,2004cs}
\label{sec:spec}

We observed SN\,2004cs with the Low Resolution Imaging Spectrometer
(LRIS; Oke \etal\ 1995) at the Cassegrain focus of the W.~M. Keck-I
10-m telescope on 2004 August 12, roughly 50 days after discovery and
45 days after peak brightness.  The supernova was quite faint at this
time so we used a nearby bright star offset, provided by the LOSS team
(Moore \& Li 2004), to place the SN on our 1\arcsec\ slit.  The
one-dimensional sky-subtracted spectrum was extracted optimally (Horne
1986) in the usual manner, with a width of 1.8\arcsec\ along the slit.
Great care was taken to properly remove the background, as the SN is
located in a complex region of its host galaxy, UGC~11001, with many
bright \ion{H}{2} regions.  The spectrum was then wavelength- and
flux-calibrated, corrected for continuum atmospheric extinction and
telluric absorption bands (Wade \& Horne 1988; Bessell 1999; Matheson
\etal\ 2000), and rebinned to 10~\AA\ pixel$^{-1}$ to improve the
signal-to-noise.  Finally, we removed a recession velocity of
4431~\kmsec\ from the spectrum, derived from the velocities indicated
by the superposed \ion{H}{2} regions.  Additional details of the
observation and reduction of this spectrum, along with the lightcurve
of SN~2004cs, will be given in a forthcoming paper (Leonard \etal\
2004, in preparation).

\begin{figure*}
\includegraphics[height=17cm, angle=90]{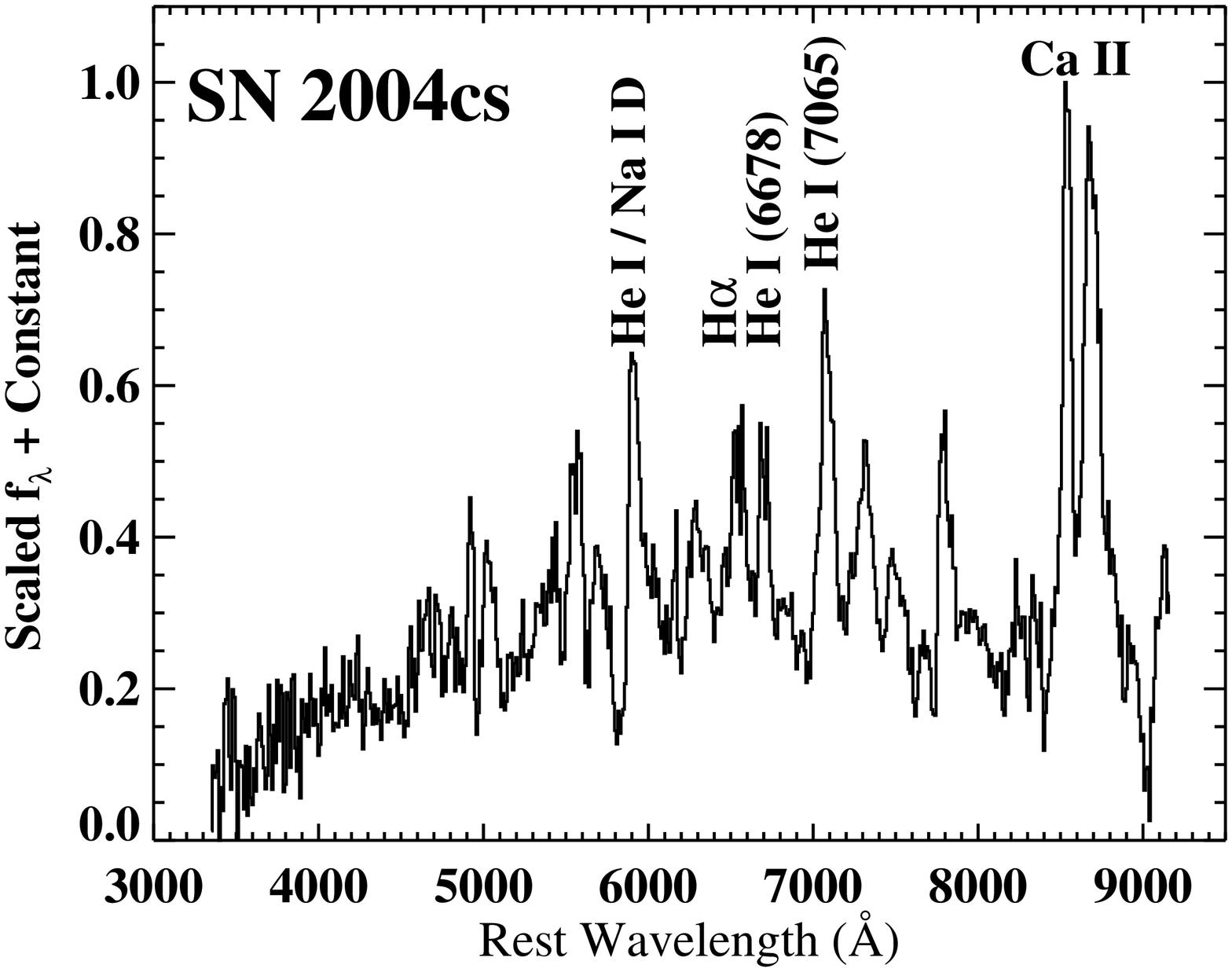}
\caption{Optical spectrum of SN\,2004cs from 2004 August 12, with
  prominent line features labeled; this date is 55~days after the
  estimated date of explosion.  A systemic velocity of 4,431~\kmsec,
  derived from the spectrum of \ion{H}{2} regions in the host galaxy,
  has been removed.  The prominent H$\alpha$ emission feature
  indicates a type~II identification, with the rarely-seen He emission
  features suggesting that SN\,2004cs is an unusual type~II, perhaps
  of subtype~IIb.}
\label{fig:sncs-spec}
\end{figure*}

We present the extracted spectrum in Figure~\ref{fig:sncs-spec}.  The
H$\alpha$ emission feature identifies SN\,2004cs as a type~II event.
The additional presence of \mbox{P-Cygni} lines of helium suggests a
type~IIb identification, although a confident determination will have
to await a joint analysis of the spectrum and light curve of this
event (Leonard \etal\ 2004).  

Although a color track for type~IIb supernovae is presented in our
photo-typing plots, such supernovae are rare, and SN\,2004cs was
discovered at an extremely young age.  Our own observations were
similarly prompt, and this combination of an unusual event and very
early observations may explain our incorrect photo-typing.  Indeed,
this shortcoming of photo-typing was known a priori; the dense
coverage of the color-color plane in Figs.~\ref{fig:sncs},
\ref{fig:sndh} and \ref{fig:sndk} means that some sort of
identification is possible for almost every event, even for unusual
SNe that do not adhere to previously-identified spectrophotometric
regimes.  In short, although our photo-typing result for this event is
incorrect, it is nonetheless incorrect in an interesting fashion --
misidentifying an unusual SN which presented us with unprecedented
colors (for a type~II SN) at our first epoch of observation.


\section{Conclusions}
\label{sec:conclude}

We have demonstrated the application of Poznanski-Gal-Yam method
(``PGM'') photo-typing (Poznanski \etal\ 2002; Gal-Yam \etal\ 2004a)
to three young supernovae observed with the Robotic 60-inch Telescope
at Palomar Observatory (P60).  This represents the first real-world
application of PGM to the problem of typing young supernovae soon
after discovery, and (for the case of SN\,2004cs) only the second
blind application of the PGM of which we are aware\footnote{During the
course of our project, a PGM photo-typing analysis of the
newly-discovered SN\,2004dj (Nakano \& Itagaki 2004) was made by
E.~Ofek and collaborators at Wise Observatory.  This analysis
correctly identified the type (II) and age (roughly one month) of the
event prior to the first spectroscopic reports (Patat \etal\ 2004a).}.

For SN\,2004cs, discovered before peak by Li \& Singer (2004), our
results provided the first constraint on the SN type, suggesting that
it was likely to be a young type~Ia event (Rajala \etal\ 2004a,b,c;
see Fig.~\ref{fig:sncs}).  To the contrary, our subsequent
\mbox{Keck-I} + LRIS spectroscopic observations
(Fig.~\ref{fig:sncs-spec}) indicate that SN\,2004cs is an unusual
type~II supernova, perhaps of subtype~IIb.  

For SN\,2004dh, discovered close to peak by Moore \& Li (2004), our
PGM photo-typing analysis suggests a type~II identification
(Fig.~\ref{fig:sndh}), consistent with the type~II spectroscopic
determination reported by Matheson \etal\ (2004).  For SN\,2004dk,
discovered before peak by Graham \& Li (2004), our PGM photo-typing
analysis suggests a core-collapse identification of either type~Ib,
type~Ic, or type~II (Fig.~\ref{fig:sndk}), consistent with the type~Ib
spectroscopic determination reported by Filippenko \etal\ (2004).

We have described in detail our approach to implementing PGM
photo-typing with multi-band optical images from a small (1.5-m)
robotic observatory.  At the first available opportunity, we image the
SN, and photometric standard fields from the Stetson catalog, in four
bands and at least two epochs.  PSF photometry with the DAOPHOT
package allows us to isolate the SN light from that of its host
galaxy, and imaging the SN field at two epochs during the night, or on
two distinct nights, allows us to control for less-than-ideal
photometric conditions.  The derived magnitudes of the SN, its
redshift as known from its host galaxy, and the constraints on its age
as reported by the SN discoverers, may then be input to the PGM
``Typing Machine'' to generate a plot comparing the SN colors with
tracks corresponding to the color evolution of SNe of various types
and appropriate ages.  These plots in turn allow an estimate of the
type of the supernova to be made.

The success of two of our three test cases suggests that PGM
photo-typing may well be useful in application to newly-discovered,
young supernovae, and that the P60 is an appropriate facility for this
work.  At the same time, our failure to correctly identify the type of
SN\,2004cs points to an inherent limitation of photo-typing, namely,
that it is unlikely to distinguish new and unusual supernovae from the
majority of more or less typical events.

In the future, we plan to distill the various aspects of our PGM
implementation into scripts which will help us to provide photo-types
for newly-discovered, young supernovae in a timely fashion.  It is our
hope that this will prove a valuable service for the larger supernova
community.

  
\section*{Acknowledgments}

The authors gratefully acknowledge the helpful comments of the
anonymous referee and the assistance of D. Poznanski, and thank
Weidong Li, Alex Filippenko, and Ryan Chornock for useful discussions
concerning SN\,2004cs.  A.~Rajala was a Mr.\ and Mrs.\ Downie D. Muir
III SURF Fellow for the duration of this research, and we acknowledge
associated administrative support from the Student-Faculty Programs
Office at Caltech, which runs the Summer Undergraduate Research
Fellowships (SURF) program.  A.~Gal-Yam acknowledges support by NASA
through Hubble Fellowship grant \#HST-HF-01158.01-A awarded by STScI,
which is operated by AURA, Inc., for NASA, under contract NAS
5-26555. D.~Leonard is supported by an NSF Astronomy and Astrophysics
Postdoctoral Fellowship under award AST-0401479.  The authors would
like to express their thanks to the research staff at Caltech and
Palomar Observatory who made the P60 automation possible.  The intial
phase of the P60 automation project was funded by a grant from the
Caltech Endowment, with additional support for this work provided by
the NSF and NASA.  The spectroscopic data presented herein were
obtained at the W.~M. Keck Observatory, which is operated as a
scientific partnership between Caltech, the University of California
and NASA, and was made possible by the generous financial support of
the W.~M. Keck Foundation.

\clearpage


\begin{deluxetable}{cccccccc}
\tabletypesize{\scriptsize}
\tablecaption{Derived Magnitudes for SNe\,2004cs, 2004dh, and 2004dk  
\label{database table}}
\tablewidth{17cm}
\tablehead{
\colhead{} & \colhead{} & \colhead{} & \multicolumn{4}{c}{Magnitudes} \\
\cline{4-8} \\
\colhead{SN} & \colhead{Date (UT)} & \colhead{Time (UT)} &  
\colhead{$B$} & \colhead{$g$} & \colhead{$V$} & \colhead{$R$} &  
\colhead{$I$}
}
\startdata
SN\,2004cs	&	24 June  
2004	&	06:59	&	18.162(75)	&	18.233(52)	&	18.060(67)	&	17.865(84)	&	 
\nodata	\\
SN\,2004dh	&	25 July 2004	&	07:58	&	18.364(44)	&	\nodata  
	&	17.906(22)	&	17.584(31)	&	17.392(38)	\\
SN\,2004dk	&	03 August  
2004	&	04:02	&	17.789(55)	&	17.500(37)	&	17.179(21)	&	16.791(17)	&	 
\nodata	\\

\enddata
\tablecomments{Positions and magnitudes of the reference stars used to  
calibrate the fields of these supernovae are available from the authors  
on request.}

\label{tab:mags}
\end{deluxetable}


\begin{deluxetable}{cccccccccc}
\tabletypesize{\scriptsize}
\tablecaption{Terms of Transformation Equations \label{database table}}
\tablewidth{17cm}
\tablehead{
\colhead{} & \colhead{} & \colhead{} & \multicolumn{3}{c}{Color Terms}  
& \colhead{} & \multicolumn{2}{c}{Airmass} & \colhead{} \\
\cline{4-6} \cline{8-9}\\
\colhead{SN} & \colhead{Filter} & \colhead{Zero-Point} &  
\colhead{$B-V$} & \colhead{$V-R$} & \colhead{$R-I$} &\colhead{} &  
\colhead{Coefficient} & \colhead{Range} & \colhead{RMS of fit}
}
\startdata
SN\,2004cs	&	B	&	0.917	&	1.031	&	\nodata	&	\nodata	&  
&	0.099	&	1.1-2.3	&	0.117 \\
&	V	&	1.719	&	$-0.022$	&	\nodata	&	\nodata	& &	0.121	&	1.1-2.4	&	0.099  
\\
&	R	&	2.124	&	\nodata	&	$-0.801$	&	\nodata	& &	0.101 	&	1.1-2.3	&	0.103  
\\

SN\,2004dh	&	B	&	1.277	&	$-0.900$	&	\nodata	&	\nodata	&  
&	\nodata	&	1.6\tablenotemark{a}	&	0.029 \\
&	V	&	2.243	&	$-0.316$	&	\nodata	&	\nodata	& &	\nodata	&	1.6	&	0.058 \\
&	R	&	2.362	&	\nodata	&	$-0.919$	&	\nodata	& &	\nodata	&	1.6	&	0.082 \\
&	I	&	2.400	&	\nodata	&	\nodata	&	$-0.969$	& &	\nodata	&	1.6	&	0.074 \\

SN\,2004dk	&	B	&	1.219	&	0.867	&	\nodata	&	\nodata	&  
&	\nodata	&	1.3\tablenotemark{b}	&	0.279 \\
&	V	&	1.473	&	0.377	&	\nodata	&	\nodata	& &	\nodata	&	1.3	&	0.183 \\
&	R	&	2.284	&	\nodata	&	$-0.808$	&	\nodata	& &	\nodata	&	1.3 	&	0.190 \\

\enddata
\tablecomments{This table denotes the coefficients of each term in the  
transformation equations used in photometric calibration.}
\tablenotetext{a}{
The SN\,2004dh calibration made use of a single standard field,  
observed at airmass 1.6.  See text for details.}
\tablenotetext{b}{
The SN\,2004dk calibration made use of a single standard field,  
observed at airmass 1.3.  See text for details.}
\label{tab:calib}
\end{deluxetable}



\end{document}